\documentclass[a4paper,12pt]{article}
\pdfoutput=1
\usepackage{jcappub}
\usepackage{tikz,xcolor,hyperref}
\usepackage{amsmath, amssymb, amsthm, graphicx, epsfig, fancyhdr,epsfig, slashed}
\usepackage[normalem]{ulem}
\usepackage{tikzsymbols}
\usepackage{natbib}
\usepackage{float}
\usepackage{bm}
\usepackage{subcaption}
\usepackage{float}
\usepackage{ dsfont }
\usepackage{comment}
\usepackage{graphicx}  
\usepackage{dcolumn} 
\usepackage{bm,relsize}   
\usepackage{slashed}
\usepackage{placeins}
\usepackage{adjustbox}
\usepackage[normalem]{ulem}
\usepackage{lscape}
\usepackage{multirow}
\newcommand{\eq}[1]{~{\rm (\ref{eq:#1})}}
\newcommand{\med}[1]{\langle #1\rangle}

\newcommand{\fig}[1]{~\ref{fig:#1}}

\def\be{\begin{equation}}
	\def\ee{\end{equation}}

\makeatletter
\font\ital=cmu10
\def\hhref#1{\href{http://arxiv.org/abs/#1}{arXiv:#1}}
\usepackage{xstring}
\newcommand{\hhrefq}[1]{\IfSubStr{#1}{:}{\href{http://inspirehep.net/search?ln=en&ln=en&p=#1&of=hb&action_search=Search&sf=&so=d&rm=&rg=25&sc=0}{InSpire:#1}}{\hhref{#1}}}

\def\art{\@ifnextchar[{\eart}{\oart}}
\def\eart[#1]#2#3#4#5#6{{\rm #2}, {\em #3 \bf #4} {\rm (#6) #5} ({\em #1})}
\def\article{\@ifnextchar[{\earticle}{\oarticle}}
\def\oarticle#1#2#3#4#5#6{{\rm #1}, {\ital `#6'}, {\rm #2 #3 (#5) #4}}
\def\earticle[#1]#2#3#4#5#6#7{{\rm #2}, {\ital `#7'}, {\rm #3 #4 (#6) #5}  [\hhrefq{#1}]}
\def\hepart[#1]#2{{\rm #2, \sl#1}}
\def\heparticle[#1]#2#3{#2, {\ital `#3'} [\hhrefq{#1}]}
\newcommand{\doi}[1]{\href{http://dx.doi.org/#1}{[link]}}

\newcommand{\hhrefqq}[1]{\IfBeginWith{#1}{10.}{\href{https://doi.org/#1}{doi:#1}}{\hhrefq{#1}}}

\newcommand{\GeV}{\,{\rm GeV}}
\newcommand{\eV}{\,{\rm eV}}
\newcommand{\TeV}{\,{\rm TeV}}

\renewcommand{\eqref}[1]{(\ref{#1})}

\newcommand{\AS}[1]{[{\color{magenta}AS: #1}]}

\definecolor{KBFIred}{RGB}{163,35,47}

\def\beq{\beq\begin{aligned}}
\def\eeq{\end{aligned}\eeq}

\def\beq{\begin{equation}\begin{aligned}}
\def\eeq{\end{aligned}\end{equation}}
\def\bea{\begin{eqnarray}}
\def\eea{\end{eqnarray}}

\begin{document}
\title{\rm\bf  Primordial magnetogenesis from a supercooled dynamical electroweak phase transition}

\author[a]{\bf Mart\'in Arteaga Tupia,}
\author[b]{Anish Ghoshal,}
\author[c]{Alessandro Strumia}

\affiliation[a]{\em Facultad de Ingenier\'ia, Universidad Privada del Norte, Lima, Per\'u
}
\affiliation[b]{\em\,\,Institute of Theoretical Physics, Faculty of Physics, University of Warsaw, Poland}
\affiliation[c]{\em\,\,Dipartimento di Fisica, Universit\`a di Pisa, Italia}
\emailAdd{martin.arteaga777@gmail.com}
\emailAdd{anish.ghoshal@fuw.edu.pl}
\emailAdd{alessandro.strumia@unipi.it}

\abstract{Observations of $\gamma$-ray from blazars suggest the presence of magnetic fields in the intergalactic medium, which may require a primordial origin.
Intense enough primordial magnetic fields can arise from theories of dynamical electroweak symmetry breaking during the big bang, 
where supercooling is ended by a strongly first order phase transition.
We consider theories involving new scalars and possibly vectors, including thermal particle dark matter candidates.
Intense enough magnetic fields can arise if the reheating temperature after the phase transition is below a few TeV.
The same dynamics also leaves testable primordial gravitational waves and possibly primordial black holes.}

\maketitle

\section{Introduction}
\label{sec:intro}

Magnetic fields observed in galaxies and galaxy clusters could have originated from plausible, though uncertain, astrophysical processes occurring after recombination. Amplification mechanisms of magnetic fields typically saturate, erasing information about the original seed fields~\cite{1303.7121,2010.10525}. 
A more reliable probe of primordial magneto-genesis is offered by coherent magnetic fields in the intergalactic medium.
Blazar observations provide indirect evidence for their existence:
blazars emit TeV-scale $\gamma$ rays; their scattering on extragalactic ambient light should produce
$e^\pm$ pairs that lead to secondary GeV-scale $\gamma$ via inverse Compton scattering on the comic microwave background.
The non-observation of such secondary $\gamma$ indicates that $e^\pm$ get deflected 
by an {intergalactic}  magnetic field. Its needed intensity
\beq  B \gtrsim 2~ 10^{-17} \,{\rm Gauss} \, \max (1, \sqrt{0.2\,{\rm Mpc}/\lambda_B}) \eeq 
depends on its unknown coherence length $\lambda_B$, possibly in the pc-Mpc range~\cite{MAGIC:2022piy, HESS:2023zwb, Neronov:2010gir}.
A stronger magnetic field is needed if blazars have longer duration, as plotted in fig.\fig{sampleB0}.

The origin of these long-correlated intergalactic magnetic fields (IGMF) is a puzzle.
Astrophysical sources
(such as the Biermann battery mechanism~\cite{PhysRev.82.863}, subsequently amplified through dynamo effects~\cite{AlvesBatista:2021sln}) 
face challenges to produce magnetic fields that fill a large volume of cosmic voids~\cite{Dolag:2010ni}.
This motivates sources originating from early universe processes. 
Possible sources of primordial magnetogenesis are inflation~\cite{Turner:1987bw, Ratra:1991bn, Martin:2007ue},
post-inflationary reheating~\cite{Kobayashi:2014sga},
some axion models~\cite{Field:2023gyx,Patel:2019isj}, cosmic strings and other defects \cite{Dimopoulos:1997df,Patel:2023sfm}, and
first-order phase transitions~\cite{Vachaspati:1991nm}.
Earlier attempts focused on the electroweak (EW) phase transition~\cite{Vachaspati:1991nm, Ellis:2019tjf,2010.10525}.
Magnetic fields arise because the weak symmetry is broken in different equivalent directions in disconnected Hubble-size patches.
The gradient in the Higgs is equivalent to electro-magnetic fields~\cite{Vachaspati:1991nm}. 
However, in the Standard Model (SM) the electro-weak phase transition is not of first order, so the resulting magnetic fields are too small.
Intense enough magnetic fields can arise in extensions of the SM where the electroweak phase transition is of first order.
Given that collider data agree with the SM, the most plausible possibility is that an extra singlet scalar undergoes a strong first order phase transition, acquiring a vacuum expectation value that affects the Higgs mass, 
indirectly inducing the weak phase transition.
The bubbles nucleated during the phase transition expand, 
collide, merge, stir the primordial plasma at high Reynolds
number, and the magnetic fields enter a regime of magnetohydrodynamic (MHD) turbulence~\cite{Witten:1984rs, Hogan:1986qda, Kamionkowski:1993fg, Brandenburg:1996fc, Christensson:2000sp, Banerjee:2004df,Kahniashvili:2010gp, Brandenburg:2017neh,2402.05179}. 
A maximal effect arises in a class of motivated models, where the whole Higgs mass and all mass scales are dynamically induced.

During the big bang, the coherence length $\lambda_B$ of such magnetic fields grows because of the universe expansion, and because of magnetic dynamics. 
In particular, magnetic helicity is approximately conserved and affects the evolution of magnetic turbulence over time, 
transferring the energy from smaller to larger length scales~\cite{Kahniashvili:2012uj, Copi:2008he}, 
and, thus, forming coherent magnetic structures at scales much larger than the ones where the energy was initially injected. 
A significant magnetic helicity can arise when parity P
and CP are substantially violated~\cite{Forbes:2000gr,Tashiro:2013ita, Chen:2014qva,Brandenburg:2017neh}.

\medskip

After summarising in section~\ref{sec:mag} how primordial magnetic fields are approximated, 
we explore three distinct scenarios of dynamical electroweak symmetry breaking, all originating from a classically scale-invariant potential. Each model features a Standard Model singlet `dilaton' field, $s$, whose self-quartic coupling becomes negative at low energies due to additional interactions: a quartic coupling with another singlet $s'$ in section~\ref{sec:mods'}, a dark SU(2) gauge interaction in section~\ref{sec:SU2}, and a U(1)$_{B-L}$ gauge interaction in section~\ref{sec:U1}. 
In these models, thermal corrections to the potential dominate at small field values, creating a potential barrier between the origin $s = 0$ and the true vacuum. As a result, the universe undergoes significant supercooling until the vacuum energy is released through bubble nucleation. 
The subsequent bubble collisions can generate magnetic fields, gravitational waves, and potentially primordial black holes. 
Conclusions and a summary of results are presented in section~\ref{sec:concl}.

\section{Magnetic fields from cosmological phase transitions}\label{sec:mag}
Computing the magnetic fields produced during first order phase transitions requires difficult
simulations and analytic understanding.
Given that significant uncertainties remain 
we here summarise the approximation we will employ.

\subsection{Magnetic fields at production}
The phase transitions arising from dynamical symmetry breaking can be
approximated using the standard dimensionless parameters $\alpha$ and $\beta$, 
which respectively quantify the released vacuum energy relative to the radiation energy density, 
$\Delta V = \alpha \rho_{\rm rad}$, and the inverse duration of the transition, $H/\beta$. 
A strongly supercooled phase transition corresponds to $\alpha \gg 1$. 
We denote by $a_*$ the cosmological scale factor at the time of reheating, immediately following the phase transition. 
The energy density $\rho_B \sim B^2/2$ in magnetic fields $B$ generated around the time $t_*$ of the transition is estimated as 
\beq \label{eq:rhoB*}\rho_{B*} \approx \frac{ \epsilon \kappa \alpha}{1+\alpha} \Delta V_H\eeq
where $\Delta V_H = m_h^4/16\lambda_H$ with $m_h \approx 125\GeV$ and $\lambda_H \approx 0.126$ is the energy released by the Higgs {part of the} potential only,
$V_H = - m_h^2 h^2/4 + \lambda_H h^4/4$;\footnote{In theories with an extra singlet scalar $s$, the total
energy released can be decomposed as $\Delta V =\Delta V_s+\Delta V_H$.
The energy $\Delta V_s$ released from the singlet sector does not source visible magnetic fields. 
In some models, the singlet is charged under a hidden gauge symmetry, in which case $\Delta V_s$ may generate hidden magnetic fields. 
However, even in the presence of vector kinetic mixing, the contribution of hidden magnetic fields to the visible sector remains negligible~\cite{1803.08051}.}
$\epsilon \sim 0.1$ is an efficiency factor for producing magnetic fields~\cite{1902.02751};
$\kappa$   is given by a sum over physical process leading to magnetic fields.
For super-cooled phase transitions with $\alpha\gg 1$, the dominant contribution comes from
bubble collisions\footnote{Another contribution
arises from the detonating expansion of an ultra-relativistic bubble 
(with velocity $v_{\rm wall}\simeq 1$ for supercooled phase transitions), that
transfers to the plasma a fraction 
\beq\kappa_{\rm exp}= \frac{\alpha (1-\kappa_{\rm col})^2}{0.73+0.083\sqrt{\alpha (1-\kappa_{\rm col})}+\alpha (1-\kappa_{\rm col})}\eeq
of the potential energy~\cite{1004.4187,Ellis:2019tjf}.
The $1-\kappa_{\rm col}$ suppression makes this contribution to $\kappa = \kappa_{\rm col}+\kappa_{\rm exp}$  negligible,
at least unless $\alpha$ is huge~\cite{Ellis:2019tjf}.
}
\beq \kappa \simeq \kappa_{\rm col} =\rho_{\rm wall}/\Delta V \simeq 1.\eeq
The spectrum of magnetic fields at production is expected to be peaked around the typical bubble size at percolation (see e.g.~\cite{1910.13125})
\begin{equation}
\lambda_*=\frac{(8 \pi)^{1 / 3}}{\beta}\, v_{\rm wall}.
\end{equation}
At length scales $\lambda$ away from the peak, the suppression is estimated as~\cite{Brandenburg:2017rnt,2010.10525,2412.00641}\footnote{We fix a typo, 10/17 instead of 17/10, that propagated to various subsequent papers.}
\beq\label{eq:rhoBprod}
\rho_{B*}(\lambda) \approx \frac{10}{17}\rho_{B*}  
\left\{\begin{array}{ll}
({\lambda}/{\lambda_*})^{-5 } &\quad \text { for}~ \lambda \gg  \lambda_* \\
({\lambda}/{\lambda_*})^{2 / 3} &\quad \text { for} ~ \lambda \ll \lambda_*
\end{array}\right.,
\eeq
such that  the total magnetic energy density is $\rho_{B*}=\int d\ln\lambda\, \rho_{B*}(\lambda)$.  

\subsection{Magnetic fields today}
To compute the magnetic field strength today, at time $t_0$ and scale factor $a_0$, one must account for the cosmological evolution of the fields from their time of production. This evolution proceeds through two main phases: from production to recombination, and from recombination to the present. 
During the latter matter-dominated era, the magnetic field energy density simply redshifts as radiation, $\rho_B \propto 1/a^4$, so that magnetic fluxes are conserved. Meanwhile, the coherence length of the magnetic fields stretches proportionally to the scale factor, following the expansion of the universe,
\begin{equation}
    B \propto a^{- 2 } \qquad \mathrm{and} \qquad \lambda \propto a  \qquad \hbox{at $a> a_{\rm rec}$}.
\end{equation}
Instead, during the initial radiation-dominated era, magnetic fields are coupled to the cosmological plasma and undergo inverse cascades. 
Neglecting the expansion of the universe, the coherence length grows as $\lambda \propto \tau^{p_\lambda}$, 
while the magnetic field strength decreases as $B \propto \tau^{-p_B/2}$, where $\tau$ denotes the proper time.
Various estimates for the two exponents exist in the literature~\cite{physics/9903028,astro-ph/0410032,Brandenburg:2017rnt,Armua:2022rvx,2203.03573,2412.00641}, mainly depending
on whether a fraction of the magnetic fields has helical shape, as magnetic helicity
$\propto B^2 \lambda$ is conserved in a plasma with high conductivity:\footnote{Magnetic helicity density $\mathbf{A} \cdot \mathbf{B}$
where $\mathbf{B} = \nabla \times \mathbf{A}$ is a measure of the twisting, linking, and knottedness of magnetic field lines.
A helical field increases its correlation length as magnetic energy dissipates, driving an inverse cascade of energy from small to large scales.}
\begin{equation}
p_B \approx 
\begin{cases} 2/3 \cr 1 \end{cases}
\qquad \mathrm{and} \qquad
p_\lambda  \approx 
\begin{cases} 2/3 & \hbox{helical} \cr 
1/2 & \hbox{non-helical}\end{cases}.
\end{equation}
This means that even a subdominant helical component of the magnetic fields tends to become dominant at late times. 
Inverse cascades can also arise in the non-helical case, since bubble collisions and parity-violating processes during the phase transition can induce helicity in the plasma velocity field. 
A more refined approximation may be possible following~\cite{2203.03573}; 
however, we do not pursue this direction here due to potential uncertainties related to micro-stability in such estimates. Instead, we rely on the standard approximations. 
Taking into account the expansion of the universe --- where the scale factor evolves as $a \propto \tau$ during the radiation-dominated era --- yields
\begin{equation}
    B \propto a^{-2 - p_B/2 } \qquad \mathrm{and} \qquad \lambda \propto a^{1+p_\lambda} \qquad \hbox{at $a< a_{\rm rec}$}.
\end{equation} 
The shape of the magnetic power spectrum remains unchanged.
As a result,
the magnetic field spectrum today $B_{0}$ is given by  
\begin{eqnarray}\label{eq.B0}
 B_0(\lambda) =  \left(\frac{a_{\mathrm{*}}}{a_{\mathrm{rec}}}\right)^{p_B / 2}\left(\frac{a_{\mathrm{*}}}{a_0}\right)^2 \sqrt{\frac{{10}}{17} \rho_{B *}} \begin{cases}\left({\lambda}/{\lambda_0}\right)^{-5 / 2} & \text { for } \quad \lambda > \lambda_0 \\
\left({\lambda}/{\lambda_0}\right)^{1 / 3} & \text { for } \quad \lambda<\lambda_0\end{cases},
\end{eqnarray}
where the coherence scale today is
\begin{equation}\label{eq:lambda0}
\lambda_0 \equiv \lambda_B\left(t_0\right)=\left(\frac{a_{\mathrm{rec}}}{a_{\mathrm{*}}}\right)^{p_\lambda}\left(\frac{a_0}{a_{\mathrm{*}}}\right) \lambda_*
=
\lambda_* H_*
\begin{cases}
0.06\,{\rm Mpc}\, (100\GeV/T_*)^{1/3} & \hbox{helical},\cr
0.6\,{\rm kpc} \,(100\GeV/T_*)^{1/2} & \hbox{non-helical.}
\end{cases}
\end{equation}
The redshift factors are
\begin{equation}
\frac{a_0}{a_{\rm rec}}\approx 1100,\qquad
\frac{a_*}{a_{\rm rec}} = \frac{T_{\rm rec}}{T_{\rm *}} \left(\frac{43/11}{g\left(T_{\text {* }}\right)}\right)^{1 / 3}
\end{equation}
where $a_{*}, a_{\text{rec}}$ and $a_{0}$ respectively are the scale factors at reheating, recombination and present time,
and
$T_{\rm rec}= a_0 T_0/a_{\rm rec}\approx 0.26\eV$.
So the peak value of the magnetic field  is
\beq \label{eq:B0lambda0}
B_0(\lambda_0) \approx \begin{cases}
8~10^{-12}\,{\rm Gauss} \,(100\GeV/T_{*})^{7/3} & \hbox{helical},\cr
8~10^{-14}\,{\rm Gauss} \, (100\GeV/T_{*})^{5/2} & \hbox{non-helical}.
\end{cases}
\eeq
In the instantaneous reheating approximation, the reheating temperature is determined by $\pi^2 g T_*^4/30 \approx\Delta V$
where $\Delta V =\Delta V_H + \cdots$ is the energy released by the full potential difference.
When the extra term is dominant, this reheating 
temperature can exceed the critical electroweak temperature, $T_{\rm cr} \approx 160\GeV$ within the bubbles (and everywhere after bubble collisions),
such that thermal effects restore the symmetric phase $\med{h}=0$.
In such a case, the electroweak phase transition will subsequently occur again, this time as a second-order transition. 
Nevertheless, the initial aborted phase electro-weak transition is sufficient to generate magnetic fields~\cite{Ellis:2019tjf}.

\medskip

\begin{figure}[t]
$$\includegraphics[width=0.7\linewidth]{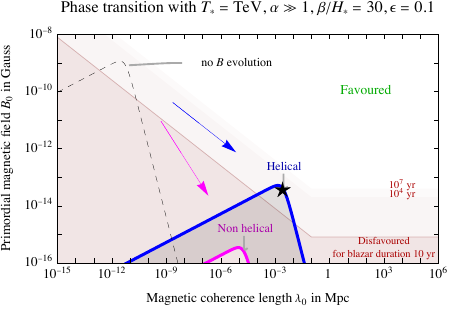}$$
 \caption{\label{fig:sampleB0}\em Sample of the predicted primordial magnetic field now $B_0$ as function of its coherence length $\lambda_0$
for a super-cooled phase transition with reheating temperature $T_*=\TeV$, percolation length $\lambda_* H_* =0.1$, efficiency $\epsilon=0.1$ for producing magnetic fields.
The blue and magenta curves include the cosmological magnetic evolution in the helical and non-helical limits, respectively,
while the dashed curve neglects such evolution.
Blazar observations suggest that intergalactic magnetic fields must lie above the shaded regions, 
depending on the uncertain duration of the blazar emission (from 10 to $10^7\,{\rm  yr}$), though the exact values of $B_0$ and $\lambda_0$
 remain unconstrained.}
\end{figure}

Fig.\fig{sampleB0} shows the primordial magnetic field generated by a supercool phase transition 
characterized by an efficiency $\epsilon=0.1$,
a reheating temperature $T_*=\TeV$ and 
an inverse duration $\beta/H_* =30$ corresponding to a characteristic scale $\lambda_* H_* = (4\pi)^{1/3} /(\beta / H_*) \approx 0.1$.
The dashed curve shows the magnetic field spectrum in the idealised limit where magnetic interactions with the plasma at $a< a_{\rm rec}$
are neglected. 
Plasma effects are incorporated in the blue and red curves, which correspond to maximal and no helicity, respectively. Since even a small initial helicity tends to grow via inverse cascade, the blue curve likely represents a more realistic outcome.
We adopt it for our subsequent plots.
Varying the assumed parameters, increasing $\lambda_*$ 
(as possibly favoured by generating Primordial Black Holes as Dark Matter) shifts the spectrum to larger $\lambda_0$,
while increasing $T_*$ reduces both $\lambda_0$ and the magnetic field $B_0$,
preventing accessing the region favoured by blazar observations if $T_* \gtrsim 3\TeV$.

\medskip

We now compute the signals in various models of dynamical symmetry breaking, focusing on those considered in~\cite{2409.04545}.
For ease of the reader we briefly summarise the relevant features of these models.

\section{Minimal model with two scalar singlets}\label{sec:mods'}
A minimal model of dynamical electroweak symmetry breaking introduces two additional scalar fields, $s$ and $s'$, in addition to the Standard Model Higgs doublet $H$. For simplicity, we assume that the theory respects the discrete symmetries $s \to -s$ and $s' \to -s'$, which constrain the scalar potential. Under these assumptions, the most general scale-invariant potential is
\beq\label{eq.Vtree}
V_{\rm tree} =V_\Lambda +\lambda_H|H|^4+\frac{\lambda_S}{4} s^4+\frac{\lambda_{S^{\prime}}}{4} s^{\prime 4}+\frac{\lambda_{H S}}{2}|H|^2 s^2  +\frac{\lambda_{H S^{\prime}}}{2}|H|^2 s^{\prime 2}+\frac{\lambda_{S S^{\prime}}}{4} s^2 s^{\prime 2} .
\eeq
The constant term $V_\Lambda$ is tuned to ensure that the potential at its minimum yields a nearly vanishing cosmological constant, i.e., $V \simeq 0$. 
The additional scalar field $s'$ plays a crucial role in enabling dynamical symmetry breaking: the quartic coupling $\lambda_{SS'}$ allows the couplings $\lambda_S$ and $\lambda_{HS}$ to run negative at lower energy scales. The specific pattern of symmetry breaking depends on which linear combination of couplings becomes negative first. Following~\cite{2204.01744,2409.04545}, we introduce a parameter $R$ to track this behaviour and determine which effective running coupling drives the breaking:
\beq
\lambda_S^{\rm eff}(s) = \frac{ \lambda_{SS'}^2}{4(4\pi)^2}  \ln \frac{s^2}{e^{1/2}w^2 },\qquad
\lambda_{HS}^{\rm eff}(s) =  \frac{\lambda_{SS'} \lambda_{HS'}}{2(4\pi)^2} \ln \frac{R s^2 }{w^2}.
\eeq
We are interested in the phase where 
$\langle s\rangle = w \gg \langle h \rangle = v=246.2$ GeV and $\med{s'}=0$.
This happens for $R\lesssim  1$ when the critical boundary 
$\lambda_{H S}=-2 \sqrt{\lambda_H \lambda_S}$ is crossed while  $\lambda_{HS}$ is negative.
In this phase $s'$  is stable.
As a result the model contains two DM candidates: the particle $s'$
and Primordial Black Holes produced during the phase transition,
from regions where bubbles appear later.
We employ the estimates from~\cite{2305.04942},
recently questioned in~\cite{2503.01962,2506.15496}.

\smallskip

\begin{figure}[t]
    \centering
    \begin{subfigure}[H]{0.48\linewidth}
        \centering
        \includegraphics[width=\linewidth]{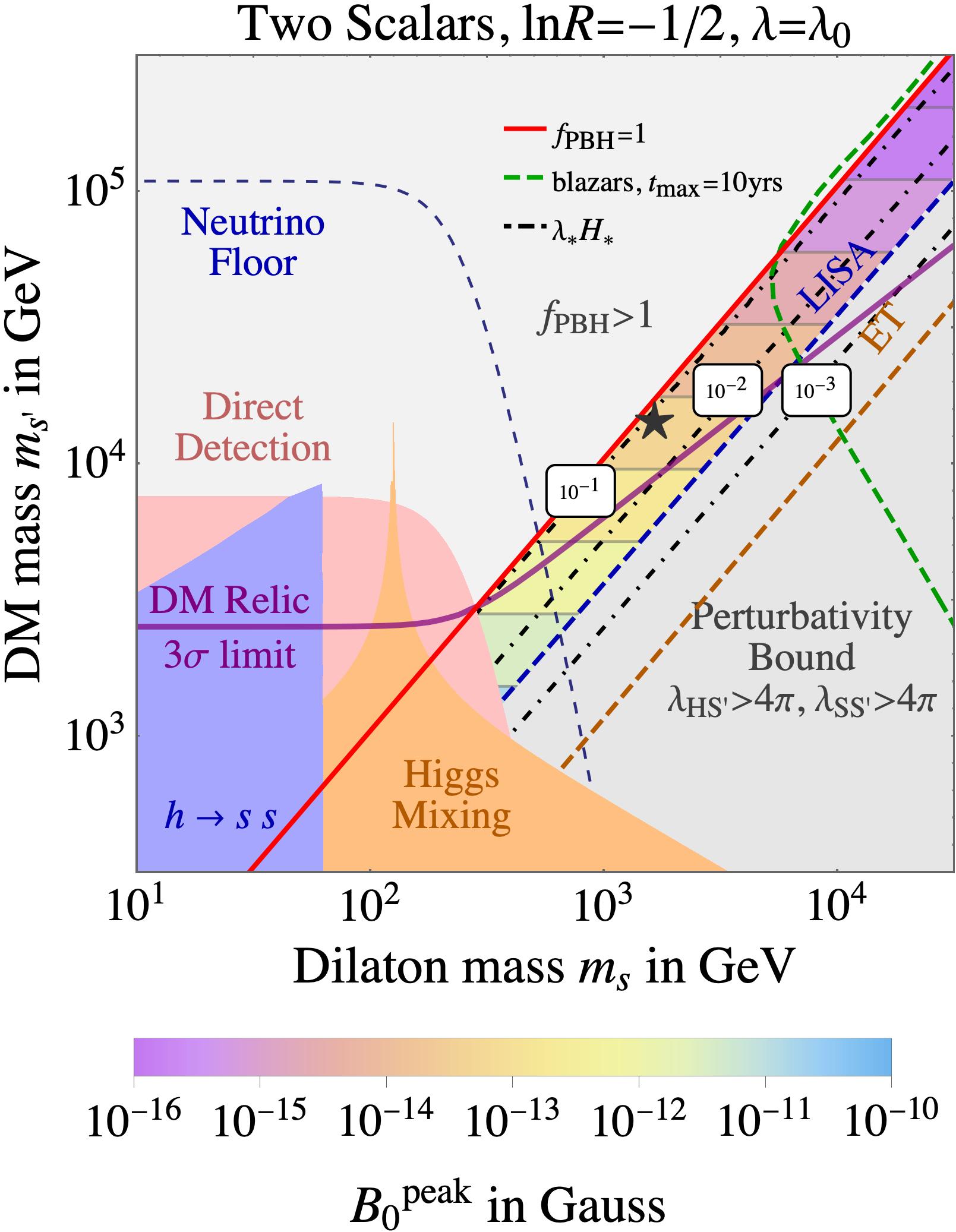}
        \label{fig:B0_1a}
    \end{subfigure}
    \hfill
    \begin{subfigure}[H]{0.48\linewidth}
        \centering
        \includegraphics[width=\linewidth]{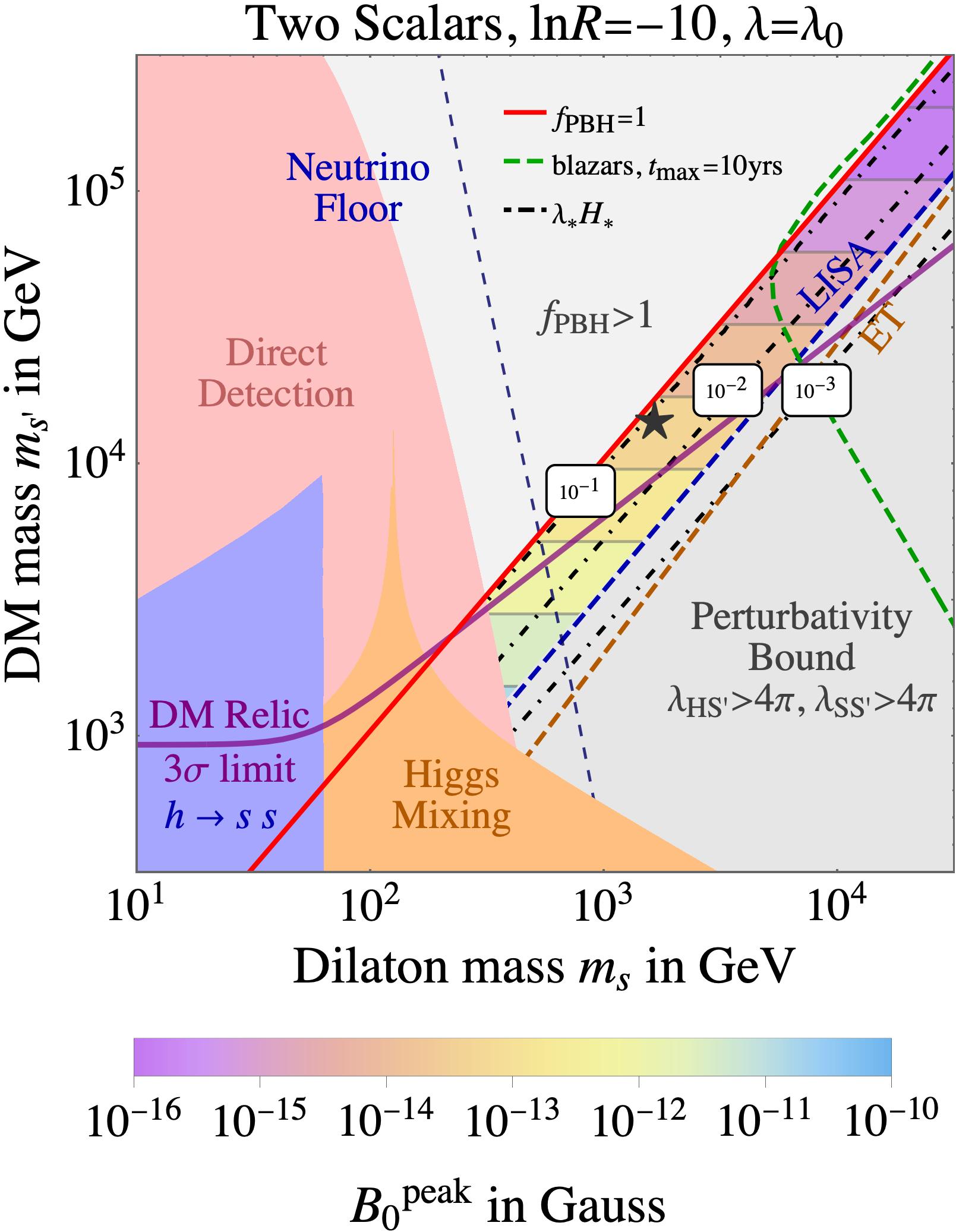}
        \label{fig:B0_1b}
    \end{subfigure}
\caption{ \em The colored contours show values of the peak value $B_0(\lambda_0)$ of the primordial magnetic field
generated by the first order phase transition. 
The dot-dashed contours show the values of the peak scale $\lambda_* H_*$, that determines $\lambda_0$ as in eq.~(\ref{eq:lambda0}).
$B_0$, computed assuming helical evolution, is intense enough to explain the blazar anomalies below the dashed green curve.
The gravitational waves produced by the phase transition are 
intense enough to be observed at LISA and ET in the regions above the curves denoted with the experiment names.
Direct detection experiments exclude the region shaded in red, and their sensitivity will reach the `Neutrino Floor' curve.
The magnetic field spectrum of fig.\fig{sampleB0} is reproduced at the point denoted with a $\star$.
 \label{fig:ss'}}
\end{figure}

The model has three free parameters, that can be chosen to be the masses of $s$ and $s'$ and $R$, fixing the other parameters to be~\cite{2204.01744}
\beq\label{eq:ss'params}
  \lambda_{SS'} &\approx \frac{(4 \pi)^{2} m_{s}^{2}}{m_{s'}^{2}},\qquad
  \lambda_{HS'} &\approx -\frac{(4 \pi)^{2} m_{h}^{2}}{m_{s'}^{2} \ln R},\qquad
  w\simeq \frac{\sqrt{2} m^2_{s'}}{4\pi m_s}.
\eeq
We show in in fig.\fig{ss'} 
two plots at fixed $R$ in the ($m_s, m_{s'}$) plane.
The relic $s'$ abundance matches the DM abundance along the magenta curve in fig.\fig{ss'}.
Perturbativity $\lambda_{HS',SS'}\lesssim 4\pi$ limits this region to DM lighter than about 10 TeV, while 
DM masses lighter than $\approx 300\GeV$ are excluded by direct detection and other collider constraints~\cite{2204.01744}.
The PBH abundance matches the cosmological DM abundance along the red curve in fig.\fig{ss'},
and exceeds it above.
Such PBH as DM curve corresponds to small $\beta/H \sim 6$, and is thereby near to the boundary where nucleation gets too slow to end supercooling.
As a result, particle $s'$ DM tends to dominate over PBH DM in the allowed region~\cite{2409.04545}.

The colored contours in fig.\fig{ss'} indicate the peak values of the present-day primordial magnetic field strength, $B_0(\lambda_0)$.
Such contours are nearly horizontal because $B_0$ is determined as in eq.\eq{B0lambda0} by the reheating temperature,
which in the current model is approximated as  $T_* \approx 0.07 m_{s'}$.
The dot-dashed contours denote the corresponding magnetic correlation length scale $\lambda_* H_*$.
Having assumed a helical magnetic field configuration, size-able values of both $B_0$ and $\lambda_0$ --- as favored by the blazar anomalies~\cite{MAGIC:2022piy, HESS:2023zwb, Neronov:2010gir} --- 
arise when the reheating temperature $T_*$ and the associated new physics scale are not far above the electroweak scale.
In contrast, higher values of $T_*$ lead to a mostly dark phase transition, suppressing the visible magnetic field signal as discussed in eq.\eq{B0lambda0}.
Fig.\fig{sampleB0} shows an example of the magnetic field spectrum. 

\smallskip

The parameter space region potentially favoured by the presence of a sufficiently large primordial magnetic field appears accessible to future experimental 
tests in multiple ways.
First, future dark matter direct detection experiments are expected to reach sensitivities approaching the so-called ``neutrino floor'', shown as a dashed curve.
Second, the stochastic gravitational wave background produced by the first-order phase transition --- computed as in~\cite{2409.04545} --- may be within the reach of future observatories such as LISA and ET. 
The regions above the dashed curves labeled `LISA' and `ET' correspond to the parameter space where these signals would be detectable
with signal-to-noise ratio (SNR) above 10.
Finally, gravitational lensing might allow detecting PBH, in case their abundance is not much below the cosmological DM abundance.

Assuming instead fully non-helical magnetic fields, the intensity drops by a few orders of magnitude, 
and the blazar anomaly can be explained only in a small portion of the parameter space.

\smallskip

The two panels of fig.\fig{ss'} with different values of $\ln R$ show similar results, given that eq.\eq{ss'params}
shows that $\ln R$ controls the $\lambda_{HS'}$ coupling relevant for DM phenomenology, but not for the phase transition. 
Qualitatively similar results will also arise in the models of dynamical symmetry breaking 
considered in the following sections.

\medskip
\section{Model with SU(2)$_{X}$ gauge interaction}\label{sec:SU2}
\begin{figure}[t]
    \centering
    \begin{subfigure}[H]{0.48\linewidth}
        \centering
        \includegraphics[width=\linewidth]{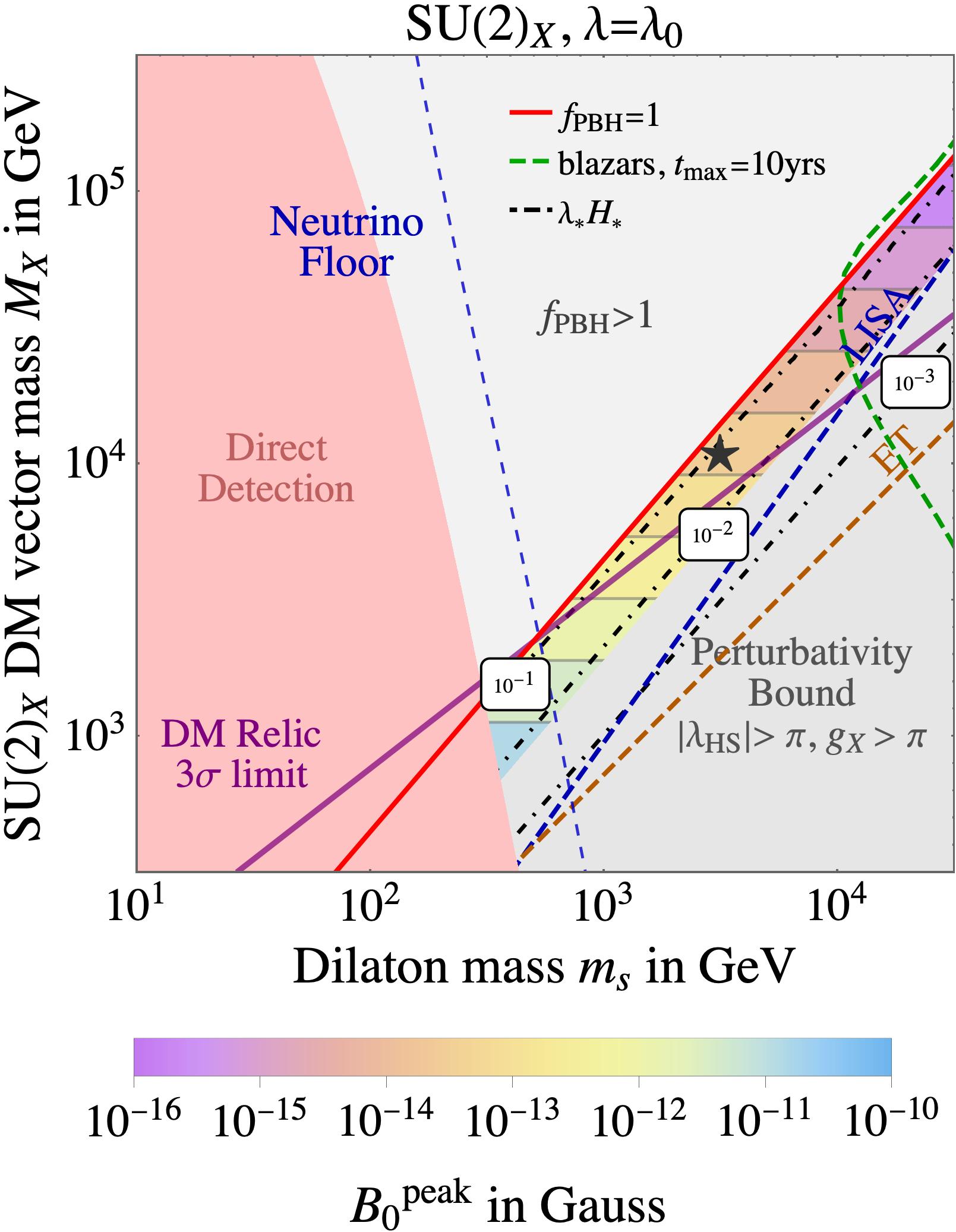}
     \end{subfigure}
    \hfill
    \begin{subfigure}[H]{0.48\linewidth}
        \centering
        \includegraphics[width=\linewidth]{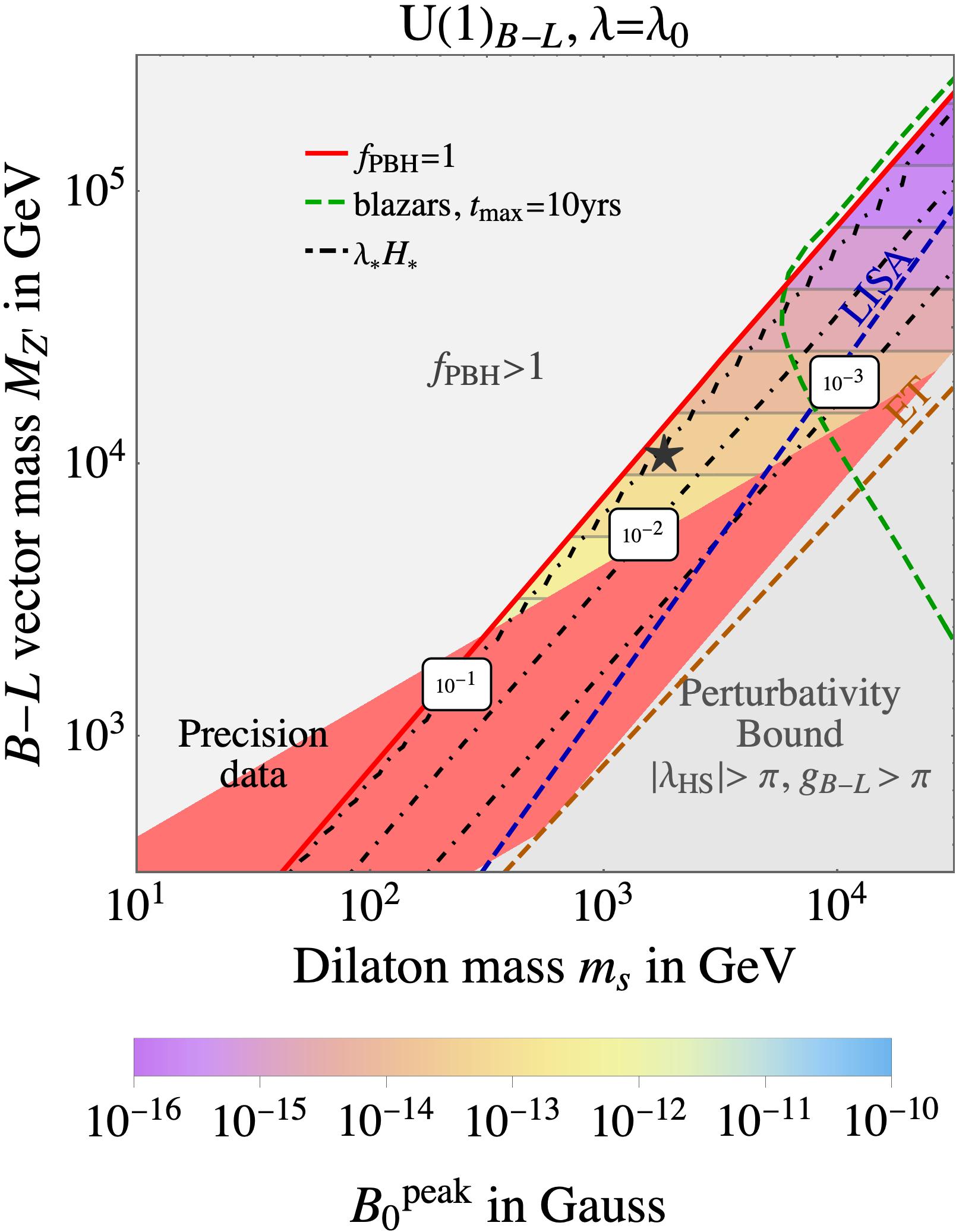}
      \end{subfigure}
\caption{\em As in fig.\fig{ss'}, for the dynamical symmetry breaking
models with ${\rm SU}(2)_X$ (left) and  ${\rm U}(1)_{B-L}$ (right) gauge interactions. 
 \label{fig:B0gauge}}
\end{figure}

A related model of dynamical symmetry  contains two scalars only: the Higgs doublet and $S$,
such that the scale-invariant scalar potential has the simpler form
\begin{equation}\label{eq:VHS}
V_{\text {tree }}=V_{\Lambda}+\lambda_H|H|^4+\lambda_S|S|^4+\lambda_{H S}|H S|^2.
\end{equation}
In these models the quartics of the scalar $S$ run negative at low energy
because $S$ is assumed to be charged under some gauge interaction.
We here assume that $S$ is a doublet under an extra SU(2)$_{X}$ with gauge coupling $g_X$,
such that $\beta_{\lambda_S} \simeq {9 g_X^4}/{8}(4\pi)^2$.
After dynamical symmetry breaking 
\begin{equation}
S=\frac{1}{\sqrt{2}}\binom{0}{s}, \quad H=\frac{1}{\sqrt{2}}\binom{0}{h}.
\end{equation}
the SU(2)$_{X}$  vectors acquire a mass $M_X = g_X w/2$
and provide a stable dark matter candidate. 
As in our previous calculation, we can assume that the free parameters of the theory are
the dilaton mass $m_s$ and the DM mass $M_X$, and write the dependent parameters as~\cite{2409.04545}
\begin{equation}
g_X \simeq \frac{4 \pi \sqrt{2} m_s}{3 M_X}, \quad \lambda_{H S} \simeq-\frac{8 \pi^2 m_h^2 m_s^2}{9 M_X^4}, \quad w \simeq \frac{3 M_X^2}{2 \sqrt{2} \pi m_s}.
\end{equation}
Our results are shown in the left panel of fig.\fig{B0gauge}. 
Both its notations and its results are similar to the previous model in section~\ref{sec:mods'},
as only order unity factors differ in the key processes.
In particular, the reheating temperature is now approximated as $T_* \approx 0.12 M_X$.

\section{Model with U(1)$_{B-L}$ gauge interaction}\label{sec:U1}
A related model assumes that a complex singlet scalar $S$ with charge $q_S$ under gauged U(1)$_{B-L}$ with gauge coupling $g_{B-L}$. 
Here $B$ is baryon number, $L$ is lepton number, and $B-L$ is their anomaly-free combination.
The scalar potential is as in eq.\eq{VHS}.
The beta function $\beta_{\lambda_S} \simeq {6(q_S g_{B-L})^4}{/(4 \pi)^2}$ of
the quartic $\lambda_S$ induces dynamical symmetry breaking at low energy, such that the scalars are expanded as
\begin{equation}
S=\frac{s}{\sqrt{2}}, \qquad H=\frac{1}{\sqrt{2}}\binom{0}{h}.
\end{equation}
We  again use $m_s$ and $M_{Z'}$ as the free parameters, where $M_{Z'}=q_S g_{B-L} w$ is the mass of the $B-L$ vector.
In the current model $Z'$ decays into quarks and leptons and thereby cannot be stable DM.
The gauge coupling, the mixing quartic scalar coupling $\lambda_{HS}$ and the vacuum expectation value $w$ can be written as
\begin{equation} \label{eqmass}
q_S g_{B-L} \simeq \sqrt{\frac{2}{3}} \frac{2 \pi m_s}{M_{Z^{\prime}}}, \qquad \lambda_{H S} \simeq-\frac{8 \pi^2 m_h^2 m_s^2}{3 M_{Z^{\prime}}^4}, \qquad w \simeq \sqrt{\frac{3}{2}} \frac{M_{Z^{\prime}}^2}{2 \pi m_s} .
\end{equation}
Results, shown in the right panel of fig.\fig{B0gauge}, are qualitatively similar to the previous models up to a main difference.
The current model does not contain a particle DM candidate, such that the region with larger $B_0$ at weak-scale masses is not excluded by direct detection bounds. 
Nevertheless, the region shaded in red is excluded by electroweak precision data, that set the bound $M_{Z'}/g_{B-L}  \gtrsim 7\TeV$.
Since $T_* \approx 0.09 M_{Z'}$, these bounds exclude the region with low $T_* $ where $B_0$ would be maximal. 

\smallskip

The scale-invariant $B-L$ model has been already studied in~\cite{Ellis:2019tjf,2402.05179} at a few benchmark points.
Our results agree within a factor of 2 (see e.g.\ footnote 3)
with the benchmark points used there-in, 
as we assumed the same approximations for the generation of primordial magnetic fields and their cosmological evolution.
The agreement holds despite a difference in notations: our efficiency factor $\kappa=1$ corresponds to their $\kappa_h =\Delta V_H/\Delta V$.
Unlike us, these authors also present a second, different,  approximation where the full $\Delta V$ (rather than just the
smaller jump $\Delta V_H$ in the Higgs potential energy) contributes to the magnetic energy density in eq.\eq{rhoB*}.
Under this different approximation, that corresponds to $\kappa_h =1$ in their notation, the authors of~\cite{Ellis:2019tjf,2402.05179}
obtain large magnetic fields even if $M_{Z'}\gg M_Z$.
We do not adopt this scenario, as we regard it as unrealistically optimistic, even in the $B-L$ model~\cite{1803.08051}.

\medskip

\section{Conclusions}\label{sec:concl}
We studied the generation of intergalactic magnetic fields from strongly first-order phase transitions in the early universe, arising in models of dynamical electroweak symmetry breaking. These scenarios extend the Standard Model by introducing an additional scalar field $s$, 
neutral under SM gauge interactions. The weak scale emerges dynamically when $s$ acquires a vacuum expectation value after a period of significant supercooling, terminated by a first-order phase transition capable of sourcing primordial magnetic fields. 

We analyzed three representative models that realise dynamical symmetry breaking from a classically scale-invariant potential:
a minimal model with a second scalar singlet; a model with a hidden SU(2) gauge interaction, a model with a gauged U(1)$_{B-L}$.
These interactions induce radiative corrections that induce dynamical symmetry breaking.
For each model, we computed the spectrum of magnetic fields generated during bubble collisions and followed their evolution to the present day, incorporating magnetohydrodynamic effects such as inverse cascades and the role of magnetic helicity. 
The coherence length  of primordial magnetic fields grows in view of the universe expansion and of
inverse cascade in the radiation-dominated plasma.
This growth depends on the helicity of the initial magnetic field and plasma velocity.
As a result, these models can produce magnetic field strengths and coherence lengths consistent with observational hints from blazar-induced $\gamma$-ray spectra, potentially explaining the origin of the inter-galactic magnetic fields. The parameter regions that yield sufficiently intense magnetic fields are also predictive in other ways. 

The same first-order phase transitions responsible for magneto-genesis can also generate a stochastic gravitational wave background within reach of upcoming detectors such as LISA and the Einstein Telescope. In addition, these transitions can lead to the formation of primordial black holes (PBHs) in a mass range where they may account for all or part of the dark matter. Mergers of such PBHs produce gravitational wave signals~\cite{1707.01480}. Some models further include thermal particle dark matter candidates with weak-scale masses and the correct relic abundance from freeze-out, adding another testable signature.
In conclusion, these scenarios appear cosmologically testable thanks to their multi-messenger phenomenology.

\small

\paragraph{Acknowledgements.} 
We thank Kohei Kamada, Marek Lewicki,  Yuhsin Tsai, Tanmay Vachaspati for clarifications.


\end{document}